%% file: LabConstrictor.tex
\def\rxivcitationstyle{author-date}

\documentclass[times, twoside, watermark]{rxiv_maker_style}
\usepackage{blindtext} 
\usepackage{float} 

\renewcommand{\today}{2026-03-11}

\leadauthor{Hidalgo-Cenalmor}

\begin{document}
\title{Packaging Jupyter notebooks as installable desktop apps using LabConstrictor}

\shorttitle{LabConstrictor}

\author[1,2]{Iván Hidalgo-Cenalmor}
\author[1,2,3]{Marcela Xiomara Rivera Pineda}
\author[4]{Bruno M. Saraiva}
\author[4,5]{Ricardo Henriques}
\author[1,2,3,6,\Letter]{Guillaume Jacquemet}
\affil[1]{Turku Bioscience Centre, University of Turku and Åbo Akademi University, Turku, FI}
\affil[2]{Faculty of Science and Engineering, Cell Biology, Åbo Akademi University, Turku, FI}
\affil[3]{InFLAMES Research Flagship Centre, University of Turku, Turku, FI}
\affil[4]{Instituto de Tecnologia Química e Biológica António Xavier, Universidade Nova de Lisboa, Oeiras, Portugal}
\affil[5]{UCL Laboratory for Molecular Cell Biology, University College London, London, United Kingdom}
\affil[6]{Foundation for the Finnish Cancer Institute, Tukholmankatu 8, Helsinki, FI}

\maketitle

\begin{abstract}

Life sciences research depends heavily on open-source academic software, yet many tools remain underused due to practical barriers. These include installation requirements that hinder adoption and limited developer resources for software distribution and long-term maintenance. Jupyter notebooks are popular because they combine code, documentation, and results into a single executable document, enabling quick method development. However, notebooks are often fragile due to reproducibility issues in coding environments, and sharing them, especially for local execution, does not ensure others can run them successfully. LabConstrictor closes this deployment gap by bringing CI/CD-style automation to academic developers without needing DevOps expertise. Its GitHub-based pipeline checks environments and packages notebooks into one-click installable desktop applications. After installation, users access a unified start page with documentation, links to the packaged notebooks, and version checks. Code cells can be hidden by default, and run-cell controls combined with widgets provide an app-like experience. By simplifying the distribution, installation, and sharing of open-source software, LabConstrictor allows faster access to new computational methods and promotes routine reuse across labs.

\end{abstract}


\begin{corrauthor}
(G. Jacquemet)
\end{corrauthor}

\section*{Introduction}

Open-source academic software has become essential to life sciences research. It facilitates the collection, storage, analysis, visualisation, and sharing of data. Microscopy is a key example: modern imaging routinely produces large, complex datasets, including 3D volumes, multiplexed acquisitions, and time-lapse movies. The amount of information that can be extracted from these data continues to grow as image analysis methods advance. Extracting quantitative biological insights from data involves navigating a broad, rapidly evolving software ecosystem, ranging from well-known graphical applications such as Fiji \citep{schindelin2012} or QuPath \citep{bankhead2017} to research code available as Python toolboxes \citep{weigert2018,stringer2021,schmidt2018}.

State-of-the-art image analysis algorithms, especially AI-driven methods, are often implemented in Python because it allows quick iteration and easy reuse of community libraries. However, Python-based tools can be hard for non-experts to use in practice because they often require installing and managing complex dependencies across different operating systems. At the same time, many academic developers lack the resources to package, distribute, and maintain user-friendly software, which widens the gap between rapid method development and everyday real-world application. Therefore, the impact of software depends not only on creating better algorithms but also on making sure others can easily install it and that it continues to work over time.

\begin{figure}[!htbp]
\centering
\makebox[\linewidth][c]{
  \includegraphics[width=\linewidth,keepaspectratio,draft=false]{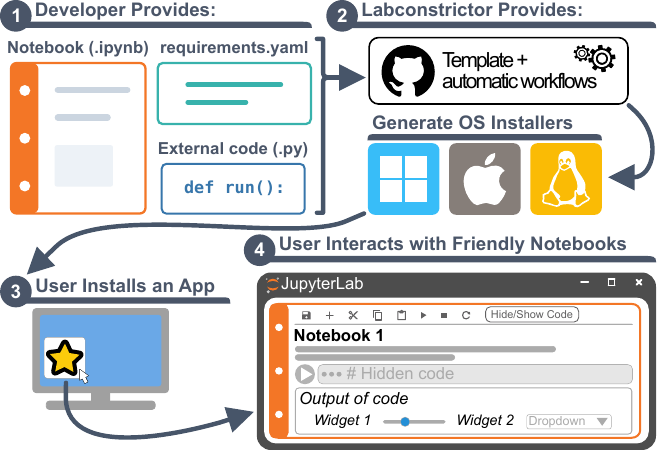}
}
\begingroup\captionsetup{width=\linewidth,singlelinecheck=false,justification=justified}\setlength{\abovecaptionskip}{6pt}\setlength{\belowcaptionskip}{6pt}\ifdefined\justifying\justifying\fi\caption{\textbf{LabConstrictor overview.} High-level schema showing how developer inputs (Jupyter notebooks, optional external Python code, and dependency specifications) populate a GitHub template repository to produce OS-specific installers, a desktop application, and a guided notebook experience in JupyterLab.}\label{fig:Figure1}\endgroup
\end{figure}

Jupyter notebooks offer a practical balance between rapid development and dissemination. By combining code, narrative explanations, and outputs within a single executable document, notebooks facilitate the presentation of the full analysis workflow and support more guided interaction \citep{perkel2018}. Tools such as ipywidgets, EZInput \citep{saraiva2026}, and marimo enable developers to expose parameters and controls, thereby minimising the need for users to modify code directly. Consequently, notebooks have become a common format for sharing computational approaches in data science and increasingly in the life sciences. However, “sharing a notebook” does not necessarily mean it is reusable: a notebook that functions for its author might not work elsewhere due to differences in operating systems, Python versions, or library dependencies \citep{rule2019,pimentel2019,samuel2024}.

Cloud platforms such as Binder \citep{jupyter2018} and Google Colab lower initial barriers by providing ready-to-use environments in the browser. Tools built on Google Colab, including ZeroCostDL4Mic \citep{vonchamier2021}, CellTracksColab \citep{gomezdemariscal2024}, and ColabFold \citep{mirdita2022}, have supported many research projects. However, cloud execution is not always feasible when clinical data and other sensitive microscopy datasets cannot leave institutional firewalls due to privacy, governance, or compliance requirements. In addition, many imaging datasets are simply too large and complex to upload to cloud resources (e.g., 3D volumes or entire pathology slides). In these cases, researchers still require reliable methods for running notebook-based workflows reproducibly on local or institutional infrastructure.

Several efforts aim to enhance the local reproducibility and accessibility of notebook workflows, including DL4MicEverywhere \citep{hidalgocenalmor2024}, JupyterLab Desktop, and album \citep{albrecht2021}. DL4MicEverywhere packages notebooks using Docker to improve stability and portability, but Docker can introduce setup and operational complexity (e.g., resource management and GPU configuration) and is sometimes constrained by institutional IT and security policies. JupyterLab Desktop makes it easy to launch notebooks, but users still need to manage Python environments. Album supports cross-platform sharing of scientific software, yet preparing tools for distribution can still require significant developer time and familiarity with packaging conventions.

Here, we present LabConstrictor, a framework that reduces practical barriers by packaging and distributing Jupyter notebooks as installable desktop applications (Fig. \ref{fig:Figure1}). For developers, LabConstrictor offers a well-documented, zero-command-line workflow built around Continuous Integration (CI) and Continuous Deployment (CD) automation for validation, packaging, and releases (Fig. \ref{fig:Figure2}{}A). For end users, it delivers an installation and launch experience that feels more like using a standalone application than managing a Python environment (Fig. \ref{fig:Figure2}{}B).

\section*{Making notebooks easy to package and install}

\begin{figure*}[p!]
\centering
\makebox[\textwidth][c]{
  \includegraphics[width=0.950\textwidth,keepaspectratio,draft=false]{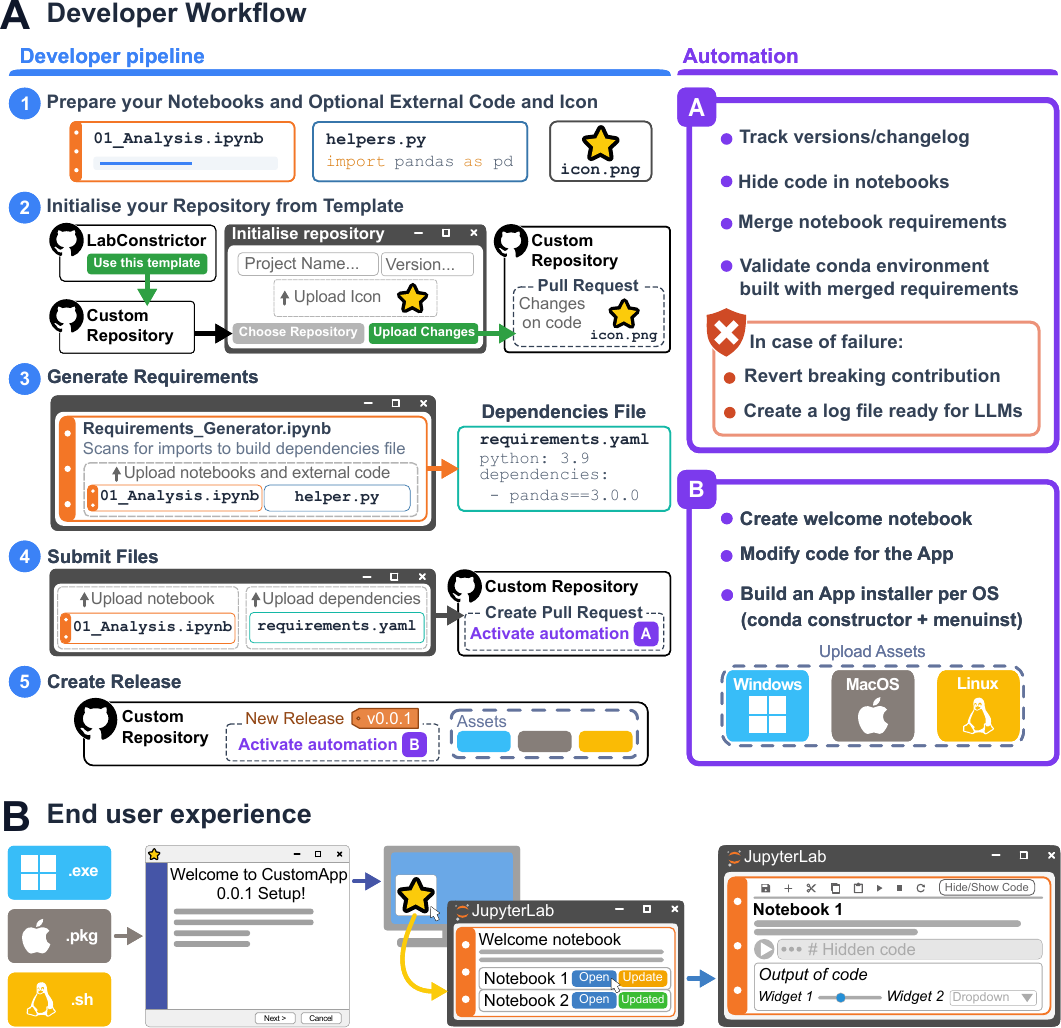}
}
\begingroup\captionsetup{width=0.95\textwidth,singlelinecheck=false,justification=justified}\setlength{\abovecaptionskip}{6pt}\setlength{\belowcaptionskip}{6pt}\ifdefined\justifying\justifying\fi\caption{\textbf{LabConstrictor workflow for developers and end-users.} (\textbf{A}) Authors prepare notebooks (and optional assets), initialize a custom repository from the LabConstrictor template, generate or curate requirements, and submit notebooks and dependencies through web forms (steps 1–5). GitHub Actions automate validation, notebook formatting, and environment setup (step A), and handle release packaging and generate installers for Windows, macOS, and Linux (step B). (\textbf{B}) End users install the desktop app via a standard installer, launch a local JupyterLab session with a welcome notebook for navigation, and run notebooks with reduced code exposure and simple execution controls.}\label{fig:Figure2}\endgroup
\end{figure*}

LabConstrictor is designed for notebook authors who want to distribute a functional workflow without converting it into a full software project with a dedicated graphical interface. It is shared as a GitHub template repository with a predefined structure, starter documentation, and automated GitHub Actions that manage most validation and packaging steps (see \textbf{Automation} in Fig. \ref{fig:Figure2}{}A). Developers mainly interact with the setup and packaging processes via the GitHub web interface and a web-based configuration form. These interfaces are used to set repository-specific options (including basic branding), manage notebooks and their dependencies, and create releases with the corresponding executables (see \textbf{Developer pipeline} in Fig. \ref{fig:Figure2}{}A).

Converting a notebook into an installable app requires minimal direct code editing and follows simple, guided steps. Additionally, to support dependency specification, which is often one of the most time-consuming parts, LabConstrictor provides a helper notebook that assesses the environment where the notebook runs (locally or online) to generate an initial requirements file, which can then be refined and validated. LabConstrictor also offers documentation and suggestions to adapt notebooks originally built for Google Colab (e.g., handling paths and environment-specific user interfaces) so they can run smoothly in a local environment. Finally, LabConstrictor enables developers to include external Python code alongside notebooks to share functionality across multiple workflows. A minimal demonstration repository derived from the LabConstrictor template is also provided at \url{https://github.com/CellMigrationLab/LabConstrictor_Demo} to illustrate the expected repository structure and workflow setup.

Because LabConstrictor is distributed as a template repository, each packaged workflow resides in its own GitHub repository, allowing developers full control over documentation, visibility, and access permissions. This is especially important for core facilities and institutional deployments where workflows may need to stay private while still benefiting from LabConstrictor’s automated validation, packaging, and local execution. 
LabConstrictor allows offline use after installation, which is essential for secure environments, institutional firewalls, and low-connectivity settings. For notebooks that do not depend on remote resources (such as repository cloning or dataset downloads), the packaged workflow operates completely locally.

\section*{Automated validation and version awareness}

As notebook workflows evolve, changes in dependencies can introduce incompatibilities that compromise functionality. LabConstrictor mitigates this risk by providing default GitHub Actions continuous integration (CI) workflows that validate the packaged environment (by creating a fresh conda environment and installing the specified dependencies) and prevent the release of non-functional deployments (see \textbf{Automation} in Fig. \ref{fig:Figure2}{}A). Once a stable environment is reached, a release-triggered CI workflow builds installers for Windows (.exe), macOS (.pkg), and Linux (.sh) using conda’s constructor, enabling a zero command-line interaction for both developers and end users.

The installers provide a user-friendly graphical interface and automatically set up the validated environment, creating a desktop application via conda’s menuinst. This simplifies the process for end users, who can install the workflow like a typical desktop application rather than building Python environments or resolving package conflicts.

To support consistent maintenance as notebooks evolve, LabConstrictor offers lightweight helper components that track versioning and notify users of available updates. Where supported, these components can also update notebooks without requiring users to reinstall the entire application. When failures occur during automated workflows, LabConstrictor generates logs intended to help with troubleshooting. These logs are designed to serve as input to large language models to facilitate faster diagnosis of common errors and, when necessary, improve communication between users and developers.

\section*{From installer to analysis: the user workflow}

For end users, LabConstrictor simplifies setup and makes it easy to find and use installed notebooks. There is no need to create Python environments, install additional packages, or resolve dependency conflicts. LabConstrictor also provides guidance through the installation process and on using notebooks in local JupyterLab sessions, making the experience more user-friendly (Fig. \ref{fig:Figure2}{}B).

End users can start by installing the app on their computers, and since it runs locally, they can access their own data and resources directly. After downloading and running the installer, users see a desktop application. When they open the app, a local JupyterLab session launches in the default browser, displaying a welcome notebook to help users get started. The goal is to make the workflow feel like using an app, rather than just running code.

The welcome notebook not only guides users but also performs version checks to ensure they are using the correct environment. It provides a clear starting point for exploring available notebooks. Within each notebook, code can be hidden, and users can reveal cells with simple controls, similar to Google Colab but running locally. 

A key benefit of LabConstrictor is its ability to run offline. For workflows that don't need remote resources (such as cloning repositories or downloading datasets), the packaged application can operate completely offline after installation. This is especially important for sensitive data and clinical research within restricted institutional environments.

\section*{Conclusions}

As AI-driven analysis in the life sciences increasingly depends on quickly evolving Python workflows, reproducible local deployment becomes essential, especially when data cannot leave institutional infrastructure. LabConstrictor supports this by making setup easier and enabling reliable local use of Jupyter-based workflows. It packages notebooks into installable desktop applications that run in a validated environment, reducing barriers for both developers and end users. In many labs, the main challenge is not creating a notebook but ensuring it works across different computers. By closing this deployment gap, LabConstrictor helps turn quickly shared notebook methods into tools regularly used in practice.

\section*{Methods}

\subsubsection{Repository initialisation and customisation}

LabConstrictor is distributed as a GitHub template repository containing a predefined folder structure, internal documentation, and automated workflows. When a developer instantiates the template, the resulting repository initially contains default placeholders. To support rapid customisation and provide a consistent end-user experience, we developed a web-based configuration form implemented as a Streamlit application (\url{https://labconstrictor-form.streamlit.app/}). The form collects basic project metadata (project name and initial version) and, optionally, images (e.g., logo, header, and welcome image) for branding.

After submission, the developer provides the GitHub repository URL and a GitHub Personal Access Token (PAT) to authorise access to the repository. The configuration form then replaces placeholders within the template repository and opens a pull request containing the customised content, allowing the developer to review and merge the changes.

\subsubsection{Dependency specification and requirements generation}

Accurately specifying dependencies for notebook workflows can be challenging, particularly when version pinning is required for reproducibility. LabConstrictor includes a requirements generation notebook to guide developers through this process. This notebook needs to be executed in the same environment where the target workflow already runs (e.g., a local conda environment/virtualenv, or a Google Colab session), so the tool can access installed packages and their versions directly.

The requirements generation notebook requests the path to the target notebook and scans code cells for import statements to infer required packages. It also detects installation commands (e.g., pip install) and prompts the developer to confirm that these packages are installed in the active environment so they can be versioned correctly. If external Python code is used (see ‘External code support’ section), the notebook can additionally scan specified .py files to capture imports outside the notebook. Detected packages are then validated against the current environment, and version information is recorded.

The output is a requirements.yaml file that includes (i) a short notebook description for display purposes, (ii) the Python version, and (iii) a versioned list of dependencies. In addition, for developers who prefer to write requirements.yaml manually, we provide a validation notebook to check the file format.

\subsubsection{Uploading notebooks and triggering automated validation}

Once notebooks and their requirements have been validated, they are uploaded to the repository using the web form (\url{https://labconstrictor-form.streamlit.app/}). This form supports uploading multiple notebooks and their corresponding requirements files, and creates a pull request that inserts files into the correct repository structure. When the pull request is merged, an automated GitHub Actions workflow is triggered to validate the contribution.

This automation handles (i) version tracking and changelog updates, (ii) formatting steps for notebooks (including code hiding, see ‘Notebook code hiding and app-like interaction’), (iii) merging requirements from submitted notebooks, and (iv) environment validation. When merging requirements, the workflow first ensures that all notebooks specify compatible Python versions; if they differ, it fails early. If Python versions are consistent, dependencies are combined into a single specification, and in case of conflicting version pins, the workflow chooses the latest version.

Environment validation runs on Ubuntu to optimize time and computational resources. A new conda environment is created using the combined Python version, and dependencies are installed from PyPI. If environment setup or package installation fails, the GitHub Actions workflow fails and automatically reverts the commit related to the notebook submission, ensuring that non-functional notebooks are not included in the installable distribution. Additionally, a troubleshooting log is generated that includes the relevant requirements, scripts, and error output, enabling developers to diagnose issues independently or with the help of a large language model (LLM), and communicate problems efficiently.

\subsubsection{External code support}

LabConstrictor supports workflows that depend on Python code outside the notebook (e.g., shared helper functions reused across multiple notebooks). Developers can upload external Python modules to the repository, and automated GitHub workflows will include these files in the distributable application. During installation, a setup.py provided by LabConstrictor installs the external code as a Python package inside the created environment, enabling standard imports from notebooks.

\subsubsection{Version tracking and update awareness}

To simplify version tracking, developers only need to define a variable (current\_version = "0.0.1") in any notebook cell. A GitHub Actions script scans notebook code cells, retrieves the specified version, and stores it in a YAML file to ensure reliable checking.

Version information is provided to end users through the welcome notebook launched by the desktop application. The welcome notebook compares the locally installed notebook version to the latest available in the repository and notifies the user if an update is ready. When supported, the interface offers an update option to download the latest notebook version without needing a full application reinstall. Additionally, the desktop application itself is versioned via repository releases, enabling update notifications when necessary.

\subsubsection{Notebook code hiding and app-like interaction}

To prevent an overwhelming interface for end users, LabConstrictor defaults to hiding code cells. While standard notebook collapsing can be accidentally undone when a user clicks into a cell, LabConstrictor uses a dedicated JupyterLab plugin (jl-hide-code, distributed via PyPI) that adds a toolbar button to toggle code visibility. When code is hidden, it stays hidden until the user explicitly makes it visible again.

To enhance app-like interaction, hidden cells can show a play-style execution button near the cell, similar to Google Colab. Code hiding can also be implemented programmatically through cell tags. In LabConstrictor, these tags are added automatically during notebook submission via GitHub Actions, so notebooks come with code hidden by default. Developers can turn off this behavior during repository initialization if end users are expected to edit code directly.

\subsubsection{Building executable installers and desktop applications}

LabConstrictor generates installable executables using the conda constructor tool, which creates platform-specific installers that set up a conda environment with predefined packages. The installer bundles additional files needed for distribution, including notebooks, optional external code, the welcome notebook, and application resources. Then, a post-install script installs the specified Python dependencies and other components required for the notebook experience (e.g., JupyterLab, ipywidgets, and the jl-hide-code plugin) into the conda environment via PyPI. 

While conda with pinned dependency versions can ensure strong reproducibility, this approach is not always flawless across different operating systems and can still be impacted by the long-term availability of upstream packages (for example, if a PyPI release is removed or becomes inaccessible). Complementary strategies based on containerization, such as DL4MicEverywhere \citep{hidalgocenalmor2024}, can address these issues by fully encapsulating the runtime environment, thus enhancing cross-system consistency and long-term reproducibility, although they require a more complex setup.

To provide a desktop entry point, LabConstrictor uses menuinst to create a desktop application that launches a JupyterLab session within the installed environment and automatically opens the welcome notebook.

Together, these steps provide a guided installation process via a graphical installer and a local desktop application that launches a controlled JupyterLab environment for navigating and running the packaged notebooks.

\vspace{1em}

\begin{manuscriptinfo}
This work is licensed under CC BY 4.0.
\end{manuscriptinfo}

\begin{code}
LabConstrictor is available at \url{https://github.com/CellMigrationLab/LabConstrictor} under an MIT License. A minimal example repository showing how to build a LabConstrictor-based workflow is available at \url{https://github.com/CellMigrationLab/LabConstrictor_Demo.}
\end{code}

\begin{contributions}
G.J.and I.H.-C. conceived the study in its initial form; I.H.-C. developed the LabConstrictor framework with input from G.J.; G.J. and M.R. provided critical feedback, testing, and guidance; G.J. acquired funding; G.J.and I.H.-C. wrote the manuscript with input from all authors.
\end{contributions}

\begin{acknowledgements}
I.H.C. is funded by the Finnish Doctoral Program Network in Artificial Intelligence, AI-DOC (decision number VN/3137/2024-OKM-6). This study received funding from the Research Council of Finland (338537, 371287, and 374180 to G.J.), the Sigrid Juselius Foundation (to G.J.), the Cancer Society of Finland (Syöpäjärjestöt; to G.J.), and the Solutions for Health strategic funding for Åbo Akademi University (to G.J.). Additionally, this research was supported by the InFLAMES Flagships Programme of the Research Council of Finland (decision numbers: 337530, 337531, 357910, and 35791). G.J. is supported by the Finnish Cancer Institute (K. Albin Johansson Professorship). This project has received funding from the European Research Council (ERC) under the European Union's Horizon 2020 research and innovation programme (grant agreement No. 101001332 to R.H.). This project also received funding from the European Union through the Horizon Europe EIC Pathfinder Open program (RT-SuperES, grant agreement No. 101099654 to R.H.). Funded by the European Union. However, the views and opinions expressed are those of the authors only. They do not necessarily reflect those of the European Union or the granting authority. Neither the European Union nor the granting authority can be held responsible for them. This work was supported by a European Molecular Biology Organization (EMBO) installation grant (EMBO-2020-IG-4734 to R.H.). Further support was provided by a Chan Zuckerberg Initiative Essential Open Source Software for Science grant (EOSS6-0000000260 to R.H. and G.J.). The project has also received funding from the "la Caixa" Foundation under the project code HR25-00453 (to R.H.).
\end{acknowledgements}

\begin{interests}
The authors declare that they have no competing or financial interests.
\end{interests}

\begin{exauthor}

\end{exauthor}

\section*{Bibliography}
\bibliography{03_REFERENCES}

\onecolumn
\newpage


\input{Supplementary.tex}

\end{document}

%% file: Supplementary.tex

%% file: 03_REFERENCES.bib
@article{rule2019,
  title        = {Ten simple rules for writing and sharing computational analyses in Jupyter Notebooks},
  author       = {Rule, Adam and Birmingham, Amanda and Zuniga, Cristal and Altintas, Ilkay and Huang, Shih-Cheng and Knight, Rob and Moshiri, Niema and Nguyen, Mai H. and Rosenthal, Sara Brin and Pérez, Fernando and Rose, Peter W.},
  journal      = {PLOS Computational Biology},
  volume       = {15},
  number       = {7},
  pages        = {e1007007},
  year         = {2019},
  publisher    = {Public Library of Science (PLoS)},
  doi          = {10.1371/journal.pcbi.1007007},
}

@article{gomezdemariscal2024,
  title        = {CellTracksColab is a platform that enables compilation, analysis, and exploration of cell tracking data},
  author       = {Gómez-de-Mariscal, Estibaliz and Grobe, Hanna and Pylvänäinen, Joanna W. and Xénard, Laura and Henriques, Ricardo and Tinevez, Jean-Yves and Jacquemet, Guillaume},
  journal      = {PLOS Biology},
  volume       = {22},
  number       = {8},
  pages        = {e3002740},
  year         = {2024},
  publisher    = {Public Library of Science (PLoS)},
  doi          = {10.1371/journal.pbio.3002740},
}

@article{samuel2024,
  title        = {Computational reproducibility of Jupyter notebooks from biomedical publications},
  author       = {Samuel, Sheeba and Mietchen, Daniel},
  journal      = {GigaScience},
  volume       = {13},
  year         = {2024},
  publisher    = {Oxford University Press (OUP)},
  doi          = {10.1093/gigascience/giad113},
}

@article{perkel2018,
  title        = {Why Jupyter is data scientists’ computational notebook of choice},
  author       = {Perkel, Jeffrey M.},
  journal      = {Nature},
  volume       = {563},
  number       = {7729},
  pages        = {145-146},
  year         = {2018},
  publisher    = {Springer Science and Business Media LLC},
  doi          = {10.1038/d41586-018-07196-1},
}

@misc{saraiva2026,
  title        = {EZInput: A Cross-Environment Python Library for Easy UI Generation in Scientific Computing},
  author       = {Saraiva, Bruno M. and Hidalgo-Cenalmor, Iván and Brito, António D. and Martínez, Damián and Shakespeare, Tayla and Jacquemet, Guillaume and Henriques, Ricardo},
  year         = {2026},
  publisher    = {arXiv},
  url          = {https://arxiv.org/abs/2601.08859},
  doi          = {10.48550/ARXIV.2601.08859},
}

@article{vonchamier2021,
  title        = {Democratising deep learning for microscopy with ZeroCostDL4Mic},
  author       = {von Chamier, Lucas and Laine, Romain F. and Jukkala, Johanna and Spahn, Christoph and Krentzel, Daniel and Nehme, Elias and Lerche, Martina and Hernández-Pérez, Sara and Mattila, Pieta K. and Karinou, Eleni and Holden, Séamus and Solak, Ahmet Can and Krull, Alexander and Buchholz, Tim-Oliver and Jones, Martin L. and Royer, Loïc A. and Leterrier, Christophe and Shechtman, Yoav and Jug, Florian and Heilemann, Mike and Jacquemet, Guillaume and Henriques, Ricardo},
  journal      = {Nature Communications},
  volume       = {12},
  number       = {1},
  year         = {2021},
  publisher    = {Springer Science and Business Media LLC},
  doi          = {10.1038/s41467-021-22518-0},
}

@article{jupyter2018,
  title        = {Binder 2.0 - Reproducible, interactive, sharable environments for science at scale},
  author       = {Jupyter, Project and Bussonnier, Matthias and Forde, Jessica and Freeman, Jeremy and Granger, Brian and Head, Tim and Holdgraf, Chris and Kelley, Kyle and Nalvarte, Gladys and Osheroff, Andrew and Pacer, M and Panda, Yuvi and Perez, Fernando and Ragan-Kelley, Benjamin and Willing, Carol},
  journal      = {Proceedings of the Python in Science Conference},
  pages        = {113-120},
  year         = {2018},
  publisher    = {SciPy},
  doi          = {10.25080/Majora-4af1f417-011},
}

@article{schindelin2012,
  title        = {Fiji: an open-source platform for biological-image analysis},
  author       = {Schindelin, Johannes and Arganda-Carreras, Ignacio and Frise, Erwin and Kaynig, Verena and Longair, Mark and Pietzsch, Tobias and Preibisch, Stephan and Rueden, Curtis and Saalfeld, Stephan and Schmid, Benjamin and Tinevez, Jean-Yves and White, Daniel James and Hartenstein, Volker and Eliceiri, Kevin and Tomancak, Pavel and Cardona, Albert},
  journal      = {Nature Methods},
  volume       = {9},
  number       = {7},
  pages        = {676-682},
  year         = {2012},
  publisher    = {Springer Science and Business Media LLC},
  doi          = {10.1038/nmeth.2019},
}

@article{weigert2018,
  title        = {Content-aware image restoration: pushing the limits of fluorescence microscopy},
  author       = {Weigert, Martin and Schmidt, Uwe and Boothe, Tobias and Müller, Andreas and Dibrov, Alexandr and Jain, Akanksha and Wilhelm, Benjamin and Schmidt, Deborah and Broaddus, Coleman and Culley, Siân and Rocha-Martins, Mauricio and Segovia-Miranda, Fabián and Norden, Caren and Henriques, Ricardo and Zerial, Marino and Solimena, Michele and Rink, Jochen and Tomancak, Pavel and Royer, Loic and Jug, Florian and Myers, Eugene W.},
  journal      = {Nature Methods},
  volume       = {15},
  number       = {12},
  pages        = {1090-1097},
  year         = {2018},
  publisher    = {Springer Science and Business Media LLC},
  doi          = {10.1038/s41592-018-0216-7},
}

@article{pimentel2019,
  title        = {A Large-Scale Study About Quality and Reproducibility of Jupyter Notebooks},
  author       = {Pimentel, Joao Felipe and Murta, Leonardo and Braganholo, Vanessa and Freire, Juliana},
  journal      = {2019 IEEE/ACM 16th International Conference on Mining Software Repositories (MSR)},
  pages        = {507-517},
  year         = {2019},
  publisher    = {IEEE},
  doi          = {10.1109/MSR.2019.00077},
}

@article{hidalgocenalmor2024,
  title        = {DL4MicEverywhere: deep learning for microscopy made flexible, shareable and reproducible},
  author       = {Hidalgo-Cenalmor, Iván and Pylvänäinen, Joanna W. and G. Ferreira, Mariana and Russell, Craig T. and Saguy, Alon and Arganda-Carreras, Ignacio and Shechtman, Yoav and Muñoz-Barrutia, Arrate and Serrano-Solano, Beatriz and Barcelo, Caterina Fuster and Pape, Constantin and Lundberg, Emma and Jug, Florian and Deschamps, Joran and Ferreira, Mariana G. and Hartley, Matthew and Seifi, Mehdi and Zulueta-Coarasa, Teresa and Galinova, Vera and Ouyang, Wei and Jacquemet, Guillaume and Henriques, Ricardo and Gómez-de-Mariscal, Estibaliz},
  journal      = {Nature Methods},
  volume       = {21},
  number       = {6},
  pages        = {925-927},
  year         = {2024},
  publisher    = {Springer Science and Business Media LLC},
  doi          = {10.1038/s41592-024-02295-6},
}

@article{stringer2021,
  title        = {Cellpose: a generalist algorithm for cellular segmentation},
  author       = {Stringer, Carsen and Wang, Tim and Michaelos, Michalis and Pachitariu, Marius},
  journal      = {Nature Methods},
  volume       = {18},
  number       = {1},
  pages        = {100-106},
  year         = {2021},
  publisher    = {Springer Science and Business Media LLC},
  doi          = {10.1038/s41592-020-01018-x},
}

@article{bankhead2017,
  title        = {QuPath: Open source software for digital pathology image analysis},
  author       = {Bankhead, Peter and Loughrey, Maurice B. and Fernández, José A. and Dombrowski, Yvonne and McArt, Darragh G. and Dunne, Philip D. and McQuaid, Stephen and Gray, Ronan T. and Murray, Liam J. and Coleman, Helen G. and James, Jacqueline A. and Salto-Tellez, Manuel and Hamilton, Peter W.},
  journal      = {Scientific Reports},
  volume       = {7},
  number       = {1},
  year         = {2017},
  publisher    = {Springer Science and Business Media LLC},
  doi          = {10.1038/s41598-017-17204-5},
}

@misc{albrecht2021,
  title        = {Album: a framework for scientific data processing with software solutions of heterogeneous tools},
  author       = {Albrecht, Jan Philipp and Schmidt, Deborah and Harrington, Kyle},
  year         = {2021},
  publisher    = {arXiv},
  url          = {https://arxiv.org/abs/2110.00601},
  doi          = {10.48550/ARXIV.2110.00601},
}

@inbook{schmidt2018,
  title        = {Cell Detection with Star-Convex Polygons},
  author       = {Schmidt, Uwe and Weigert, Martin and Broaddus, Coleman and Myers, Gene},
  journal      = {Lecture Notes in Computer Science},
  pages        = {265-273},
  year         = {2018},
  publisher    = {Springer International Publishing},
  doi          = {10.1007/978-3-030-00934-2_30},
}

@article{mirdita2022,
  title        = {ColabFold: making protein folding accessible to all},
  author       = {Mirdita, Milot and Schütze, Konstantin and Moriwaki, Yoshitaka and Heo, Lim and Ovchinnikov, Sergey and Steinegger, Martin},
  journal      = {Nature Methods},
  volume       = {19},
  number       = {6},
  pages        = {679-682},
  year         = {2022},
  publisher    = {Springer Science and Business Media LLC},
  doi          = {10.1038/s41592-022-01488-1},
}
